\documentclass[runningheads]{svjour2}
\smartqed  
\usepackage{graphicx}
\newcommand{\be}{\begin{equation}}
\newcommand{\ee}{\end{equation}}
\newcommand{\bea}{\begin{eqnarray}}

\newcommand{\eea}{\end{eqnarray}}
\journalname{General Relativity and Gravitation}
\begin{document}

\title{The nearly Newtonian regime in Non-Linear Theories of Gravity
}



\author{Thomas P.~Sotiriou
}


\institute{Thomas P.~Sotiriou \at
              SISSA-International School of Advanced Studies, via Beirut 2-4, 34014, Trieste, Italy and INFN, Sezione di Trieste \\
              \email{sotiriou@sissa.it}           
}

\date{Received: date / Revised: date}

\maketitle

\begin{abstract}
The present paper reconsiders the Newtonian limit of models of modified gravity including higher order
terms in the scalar curvature in the gravitational action. This was studied using the Palatini variational principle in [Meng X. and Wang P.: Gen.~Rel.~Grav.~{\bf 36}, 1947 (2004)] and
[Dom\'inguez A. E. and Barraco D. E.: Phys.~Rev.~D {\bf 70}, 043505 (2004)] with contradicting results.
Here a different approach is used, and problems in the previous attempts are pointed out. It is shown
that models with negative powers of the scalar curvature, like the ones used to explain the present
accelerated expansion, as well as their generalization which include positive powers, can give the correct
Newtonian limit, as long as the coefficients of these powers are reasonably small. Some consequences of the
performed analysis seem to raise doubts for the way the Newtonian limit was derived in the purely metric
approach of fourth order gravity [Dick R.: Gen.~Rel.~Grav.~{\bf 36}, 217 (2004)]. Finally, we comment on a recent paper [Olmo G. J.: Phys.~Rev.~D {\bf 72}, 083505 (2005)] in which the problem of the Newtonian limit of both the purely metric and the Palatini
formalism is discussed, using the equivalent Brans--Dicke theory, and with which our results partly disagree.
\end{abstract}

\section{Introduction}\label{sec:1}

The cosmological puzzle of the late time accelerated expansion of the
universe \cite{super,super2,torny,bennett,boom,halverson} has motivated a
lot of authors to consider models of modified gravity for its theoretical
explanation. One type of models is the so-called fourth order gravity,
where the gravitational action involves higher order terms in the scalar
curvature, $R$ (see for example \cite{carroll,noji}). This idea is not
new; it has been studied in the past for explaining early time inflation
\cite{staro}. The main difference between the two approaches is that in
order to generate early time inflation one needs positive powers of the
scalar curvature in the action, whereas to explain the present accelerated
expansion, negative powers of $R$ should be present.

In \cite{vollick} Vollick considered a further modification, by using the
action of \cite{carroll} together with the Palatini variational principle.
According to this principle, the metric and the affine connections are
considered as geometrically independent quantities. The action has to be
varied with respect to both of them to give the field equations. This
approach gives identical results to those obtained with the metric
approach, if the standard Einstein--Hilbert action is used, the variation
with respect to the metric giving the Einstein equations and the variation
with respect to the connections giving the expressions for the Christoffel
symbols \cite{wald}. However, differences arise between the results
obtained with the two approaches when a more general action is used.

The Palatini approach can be viewed as a further generalization of fourth
order gravity. However, standard general relativity does not need the
assumption that the connections are the Christoffel symbols of the metric;
this can simply be derived from the variation of the action. In that sense
the Palatini approach seems more appealing since it requires one less
assumption than the metric approach. Another important aspect is that a
gravitational action with higher order terms in $R$, treated in this
formalism, gives a set of second order pde's plus an equation involving
the connection, which is trivial to solve and interpret using a conformal
transformation \cite{vollick}. Remarkably, in vacuum, the theory reduces to standard general relativity plus a cosmological
constant. Fourth order gravity in the metric
formalism, however, leads to fourth order differential equations (hence
the name), which are very difficult to treat. Thus, also from the
practical point of view, the Palatini approach seems preferable. It has also been shown that stringent constraints
can be put on models of cosmological interest in the metric formalism, like the one discussed in \cite{carroll}, via confrontation with the solar system experiments \cite{chiba}.
  Even though this problem can be overcome by more
sophisticated models \cite{noji}, this requires significant fine tuning of
the various parameters. Additionally, Dolgov and Kawasaki have shown in
\cite{dolgov} that a violent instability appears in models like
\cite{carroll} in a weak gravitational regime within matter, when treated
within the framework of the metric formalism, but the Palatini version of
the same models is free of such instabilities (see \cite{meng3}).

Theories like those presented in \cite{carroll} or \cite{vollick} seem to explain the present
accelerated expansion \cite{carroll,meng3} but there are obscure points
about their behavior in a weak gravity regime. Having the correct
behavior in the Newtonian limit is, of course, fundamental for any
proposed theory. In \cite{dick} constraints were obtained for the models
in the metric formalism, in order for them to reproduce gravity as we know
it from experiments in weak field.  The picture in the Palatini formalism
is less clear.
 In \cite{meng2} Meng and Wang claim that all models with inverse powers
of the scalar curvature in the action give a correct Newtonian limit. On
the other hand, in \cite{barraco} it is claimed that this is not true and
that there are constraints on the form of the Lagrangian.  
During the preparation of this paper the post-Newtonian expansion of the discussed theories 
was studied in \cite{olmo,olmo2} both for the metric and the
Palatini formalism, by using the equivalent Brans--Dicke theory.
The results relevant to the Newtonian limit seem to be in agreement with those of
\cite{dick} and \cite{barraco}. The line of reasoning in \cite{olmo2} is basically no different from that used in \cite{olmo}, at
least for what concerns this study. Thus, from now on, any reference to
\cite{olmo} will actually include also \cite{olmo2}.

In the present paper, we reconsider the weak field regime in the Palatini
formalism and show that any reasonable model gives the correct Newtonian
limit, following a different approach from both \cite{meng2},
\cite{barraco} and \cite{olmo}. This is done in section \ref{sec:2}.  
Section \ref{sec:21} is dedicated to detailing the points at which our
analysis disagrees with that in \cite{olmo}. The discussion in section
\ref{sec:2} raises doubts about whether the studies of the Newtonian limit
within the metric formalism performed by Dick \cite{dick} and Olmo
\cite{olmo}, are viable. We comment on this in section \ref{sec:3}.
Section \ref{sec:4} contains conclusions.
\section{\label{sec:2}Palatini formalism}
We will consider theories of gravity coming from an action of the form
\be
\label{action}
S=\frac{1}{2\kappa}\int d^4x\sqrt{-g} f(R)+S_M,
\ee
where $f(R)$ is a polynomial including both positive and negative powers of $R$, $\kappa=8\pi G_N$, and
$S_M$ represents the matter action. The energy momentum tensor is given by
\be
T_{\mu\nu}=-\frac{2}{\sqrt{-g}}\frac{\delta S_M}{\delta g^{\mu\nu}}
\ee
In the Palatini formalism, when we vary with respect to the metric we get
\be
\label{struct1}
f'(R) R_{\mu\nu}-\frac{1}{2}f(R)g_{\mu\nu}=\kappa T_{\mu\nu},
\ee
and contracting,
\be
\label{struct}
f'(R) R-2 f(R)=\kappa T,
\ee
where the prime denotes differentiation with respect to $R$ and $T$ is the trace of the stress energy tensor.

In \cite{meng2} and \cite{barraco} the authors expand around de Sitter in order to derive the Newtonian limit. 
 We write 
\be
R=R_0+R_1,
\ee
where $R_0$ is the scalar curvature of the background de Sitter spacetime and $R_1$ is the correction to $R_0$,
including all possible terms, with $R_1/R_0$ being considered as a small quantity.
We will need to calculate $f(R_0+R_1)$ and $f'(R_0+R_1)$. The usual approached is to 
Taylor expand around $R=R_0$ and keep only the leading order terms in $R_1$ but we will show that
this cannot be done in the present context because $R_1/R_0$ is not small.

Take as an example the CDTT model \cite{carroll}, studied by Vollick \cite{vollick} in the Palatini formalism. Then
\be
\label{cdtt}
f(R)=R-\frac{a^2}{R},
\ee
and $a\sim 10^{-67}(\textrm{eV})^2\sim 10^{-53}\textrm{m}^{-2}$.
Expanding we get
\be
f(R)=f(R_0)+f'(R_0) R_1+\frac{1}{2}f''(R_0) R_1^2+\ldots
\ee
and using (\ref{cdtt}) we get
\be
\label{expans}
f(R)=f(R_0)+\left(1+\frac{a^2}{R_0^2}\right) R_1-\frac{1}{2}\frac{2 a^2}{R_0^3} R_1^2+\ldots
\ee
where now $R_0=a$. It is easy to see then that the second term on the right hand side of the above
equation is of the order of $R_1$,
whereas the third term is of the order of $R_1^2/a$. Therefore, in order to truncate the third term
one needs $R_1\gg R_1^2/a$ or 
\be
\label{cond}
a\gg R_1. 
\ee
Note that this is not any exceptional constraint. $R_0=a$ and so this is the usual condition for being able to truncate non linear terms in a Taylor expansion.

Let us now return to eq.~(\ref{struct}). For the CDTT model (eq.~(\ref{cdtt})) this gives
\be
\label{rt}
R=\frac{1}{2}\left(-\kappa T\pm\sqrt{\kappa^2 T^2+12 a^2}\right).
\ee
When discussing whether a theory has a good Newtonian limit, we are in practice checking whether the field equations reduce to a high precision to the Poisson equation, under certain assumptions: energy densities should be small enough so that there are no strong gravity effects and velocities related to the motion of the matter should be negligible compared to the velocity of light. At the same time energy densities should be high enough so that the system under investigation can be considered gravitationally bound \footnote[2]{For example, in the cosmological constant model one could consider, even on non-cosmological scales, densities low enough so that the correction coming from the cosmological constant dominates with respect to the matter density in the Poisson equation. This of course would not imply that this model does not have a correct Newtonian limit}. As a typical example of a density satisfying the above criteria we can take the mean density of the solar system, $\rho\sim 10^{-11} \textrm{gr/cm}^3$ \footnote[3]{One could imagine that this would be the density of a dust cloud which, due to some instabilities, collapsed to form the sub-structures of the solar system (sun, planets, asteroids, etc.) in a fairly Newtonian way.}.
 This estimate now implies that  
$\left|{a/\kappa T}\right|\sim 10^{-21}$, where $T\sim -\rho$ ($c$ is taken to be equal to 1).
The ``physical'' branch of the solution given in eq.~(\ref{rt}) seems to be the one with  
the plus sign in front of the square root.
In fact, given that $T<0$, this branch ensures that the matter leads to  
a standard positive curvature in a strong gravity regime. Then
\be
\label{r}
R\sim -\kappa T-3a^2/\kappa T
\ee
and $R_1\sim -\kappa T\sim \kappa \rho$. 
Thus $a/R_1\sim 10^{-21}$ and it is now evident that condition (9) does not hold for the 
typical densities related to the Newtonian limit.

Note that the situation does not improve even if we choose the  
``unphysical'' branch of eq.~(\ref{rt}) which has a minus sign in front of the  
square root. In fact in this case $R_1\sim a (3 a/\kappa T-1/3)$ so the correction to the  
background curvature is of the order $a$ and not much
smaller than that as required in order to truncate
the higher order terms in the expansion eq.~(\ref{expans}).

In \cite{barraco}, this fact was overlooked and only linear terms in $R_1$ were kept
in the expansion of $f(R)$ and $f'(R)$ around $R_0$. In \cite{meng2} even though they notice it in the final
stages of their analysis and they use it to truncate some terms, the authors do not take it into account
properly from the beginning, keeping again only first order terms (eq.~(11) of \cite{meng2} for example).

An alternative way to attack the problem of the Newtonian limit is the following.
 We already know from relevant literature \cite{vollick} that the connections are the
Christoffel symbols of the metric
\be
h_{\mu\nu}=f'(R)g_{\mu\nu}.
\ee
For the CDTT model then, and if we define $\epsilon=a^2/R^2$, eq.~(\ref{struct1}) takes the form
\be
\label{ein2}
(1+\epsilon)R_{\mu\nu}-\frac{1}{2}(1-\epsilon)R g_{\mu\nu}=\kappa T_{\mu\nu},
\ee
and
\be
\label{hg}
h_{\mu\nu}=(1+\epsilon)g_{\mu\nu}.
\ee
Due to eq.~(\ref{struct}) $\epsilon$ depends only on $T$.
Combining eqs.~(\ref{hg}) and (\ref{ein2}) we get
\be
\label{ein}
R_{\mu\nu}-\frac{1}{2}R h_{\mu\nu}+\epsilon\left(\frac{R}{1+\epsilon}h_{\mu\nu}+R_{\mu\nu}\right)=\kappa T_{\mu\nu}.
\ee
Note that up to this point no approximation or truncation was used.
We have merely expressed the left hand side of the
 structural equation of spacetime with respect to quantities depending only on the $h_{\mu\nu}$ metric, which is conformal to $g_{\mu\nu}$.
 However, using eq.~(\ref{rt}) and (\ref{r}) we see that $\epsilon \sim 10^{-42}$ if we use consider the mean density of the solar system as before and even less for higher densities. So the 
two metrics are practically indistinguishable in such cases, due to eq.~(\ref{hg}). Thus we can 
use the $h$ metric to derive the Newtonian limit.
If we assume that
\be
h_{\mu\nu}=\eta_{\mu\nu}+h^1_{\mu\nu},\qquad |h^1|\ll 1,
\ee
 then the first two terms of eq.~(\ref{ein}) will give the standard Newtonian limit and the last two 
terms will give a negligible contribution, since they are suppressed by the $\epsilon$ coefficient. 
A deviation of the order of $10^{-42}$ is far below the accuracy of any known experiment. In fact, one can consider densities several orders of magnitude lower and still get corrections which will be much below experimental accuracies.

A critical point is that here we assumed that the metric is flat plus a small correction instead of 
de Sitter plus a small correction. Note, however, that we are not claiming that we are expanding around 
the background or any corresponding maximally symmetric spacetime. We are merely asking for the matter to
account for the deviation from flatness which is the basic concept related to the Newtonian limit.
In any case, de Sitter is essentially identical to Minkowski for the densities discussed, and the important
corrections to the metric come from the local matter, not from considerations of the universe as a whole.

According to the above the CDTT Lagrangian in the Palatini formalism gives a perfectly good 
Newtonian limit. The approach can be extended to more general Lagrangians and it is safe to assume
that all $f(R)$ theories in this formalism will give a correct Newtonian limit as long as the extra terms
are suppressed by small enough coefficients. This happens because the form of eq.~(\ref{struct}) is 
such that in all such cases, $R\sim \kappa\rho$ for densities relevant to the Newtonian limit.

\section{\label{sec:21}Recovering the Newtonian Limit through the Equivalent Brans--Dicke theory}

The problem considered here has also recently been studied in \cite{olmo} using the equivalence
between $f(R)$ theories of gravity and Brans-Dicke theories. It is shown there that the Palatini formalism
of any $f(R)$ theory is equivalent to a Brans-Dicke theory of gravity with 
Brans-Dicke parameter $\omega=-3/2$ (for the metric formalism see section
\ref{sec:3}). Keeping that in mind, one can generalize known results to obtain the post-Newtonian metric.
 Olmo used the fact that, for approximately static solutions, one can drop terms involving time derivatives
to the lowest order, and the metric can be expanded about its Minkowskian value. The final relations giving the post-Newtonian limit of the field equations where
\bea
\label{olmo11}
-\frac{1}{2}\nabla^2\left[h^1_{00}-\Omega(T)\right]&=&\frac{\kappa \rho-V(\phi)}{2\phi},\\
\label{olmo21}
-\frac{1}{2}\nabla^2\left[h^1_{ij}+\delta_{ij}\Omega(T)\right]&=&\left[\frac{\kappa \rho+V(\phi)}{2\phi}\right],
\eea
where $V$ is the potential of the scalar field $\phi$ and $\Omega(T)\equiv \log[\phi/\phi_0]$.
The subscript $0$ in $\phi_0$, and in any other quantity henceforth, denotes that it is evaluated
 at $T=0$. Note at this point that the normalization by $\phi_0$ in this 
definition is not required. Olmo probably just added (subtracted) the constant $\log(\phi_0)$
inside the brackets on the left hand side of eq. (\ref{olmo11}) (eq. (\ref{olmo21})) using the fact that
it remains unchanged. Thus we are not going to use it here and we are going to refer to $\Omega(T)$ 
just as $\Omega(T)=\log[\phi]$. The solutions of eqs. (\ref{olmo11}) and (\ref{olmo21}) are
\bea
\label{olmo1}
h^1_{00}(t,x)&=&2 G \frac {M_{\odot}}{r}+\frac{V_0}{6\phi_0}r^2+\Omega(T),\\
\label{olmo2}
h^1_{ij}(t,x)&=&\left[2\gamma G \frac {M_{\odot}}{r}-\frac{V_0}{6\phi_0}r^2-\Omega(T)\right]\delta_{ij},
\eea
where $M_\odot\equiv \phi_0 \int d^3 x' \rho(t,x')/\phi$. The effective Newton's constant $G$ and 
the post-Newtonian parameter $\gamma$ are defined as
\bea
G&=&\frac{\kappa}{8\pi \phi_0}\left(1+\frac{M_V}{M_{\odot}}\right),\\
\gamma&=&\frac{M_{\odot}-M_V}{M_{\odot}+M_V},
\eea
where $M_V\equiv \kappa^{-1} \phi_0 \int d^3 x'\left[V_0/\phi_0-V(\phi)/\phi\right]$. Note that the $\kappa^2$ appearing in these 
relations in \cite{olmo} is what we denote here as just $\kappa$. Even though we agree with the 
approach followed to derive eqs. (\ref{olmo1}) and (\ref{olmo2}) and on their validity, we disagree
with the line of reasoning used by Olmo to argue that models with inverse powers of the scalar curvature
do not have a good Newtonian limit. We will demonstrate this using again the CDTT model (eq. (\ref{cdtt})).

As stated in different words in \cite{olmo}, if we define the Newtonian mass as $M_N\equiv \int d^3 x' \rho(t,x')$, the requirement for a theory to have a good Newtonian limit is that 
$G M_\odot$ is equal to
$G_N M_N$, where $N$ denotes Newtonian and $\gamma\sim 1$ up to 
very high precision. Additionally, the second term on the right hand side of both eq.~(\ref{olmo1}) and 
eq.~(\ref{olmo2}) should be negligible, since it acts as a term coming from a cosmological constant.
$\Omega(T)$ should also be small and have a negligible dependence on $T$. The above have to 
be true for the range of densities relevant to the Newtonian limit, as discussed before. 
Using the equation that define $V$ and $\phi$ found in \cite{olmo} (see also \cite{carroll}) one can 
easily show that
\bea
\label{eq1}
\phi&=&1+\frac{a^2}{R^2},\\
V(\phi)&=&2 a\kappa\sqrt{\phi-1}.
\eea
Additionally, for $T=0$, $R=\sqrt{3} a$ so, 
\bea
\label{eq2}
\phi_0&=&4/3,\\
\label{eq3}
V_0&=&2 a\kappa/\sqrt{3}.
\eea
For the densities we consider we can use the parameter $\epsilon$ defined above. Then
\be
\label{eq4}
V(\phi)=2\kappa \frac{a^2}{R}=2\kappa a\sqrt{\epsilon},
\ee
and $M_V\sim a$. It is very easy to see using eq. (\ref{eq1}), (\ref{eq2}), 
(\ref{eq3}) and (\ref{eq4}) that
\bea
G&\approx&\frac{\kappa}{8\pi \phi_0},\\
\gamma&\approx&1,
\eea
and $\phi\approx 1$ plus corrections of order $a$ or smaller, which is well above the limit of any experiment.

Additionally
\be
\Omega(T)\equiv\log[\phi]=\log\left[1+\epsilon\right]\approx\log\left[1+a^2/\kappa^2 T^2\right].
\ee
$V_0$ is of the order of $a$ which is a perfectly acceptable value and $\Omega(T)$ is negligible 
at the densities being considered and decreases even more when the density
increases. Therefore, our previous result are valid and theories including inverse powers of the scalar curvature
have a correct Newtonian limit in the Palatini formalism.

Of course this result contradicts those reported in \cite{olmo}, even though the approach followed there seems to be satisfactory.
The main reason for this problem seems to be the following. In \cite{olmo} the fact that $\Omega(T)$ should
have a mild dependence on $T$ is used to obtain a constraint for the dependence of $\phi$ on $T$ (eq.~(26) of 
\cite{olmo}). Taking a number of steps this constraint is turned into a constraint for the 
functional from of $f(R)$ (eq.~(37) of \cite{olmo}) and from that a conclusion is derived about the
its possible nonlinearity. We disagree with this line of thought. Such inequalities constrain merely
the value of the relevant quantity at the point where it is evaluated and not its true functional form. One could probably
use them to make some assumptions about the leading order term but not to exclude any terms of a different form,
as long as they are negligible with respect to the leading order for the relevant values of $R$. This, for example, is the case for the CDTT model discussed above.
Any constraint placed by the Newtonian limit has to hold over a certain
range of relevant densities (and consequently curvatures), and at not for all densities as implied in \cite{olmo}.

It should be clarified, of course, that the approach followed in \cite{olmo,olmo2} is broad enough to investigate, apart from the Newtonian limit of the models discussed, also their post-Newtonian behavior and the confrontation with the solar system experiments. Solar system tests (light deflection, Shapiro time delay, etc.) are much more sophisticatedthan the Newtonian limit of a theory and at the same time very different. They do not examine gravitationally bound systems, but are essentially vacuum tests, in which the presence of matter ({\it e.g.~}solar winds) has to be taken into account as a correction. Therefore the relevant densities can be many orders of magnitude smaller than those associated with the Newtonian limit. The approach which we followed here only considered densities relevant for the latter and our objections to \cite{olmo,olmo2} are confined to the discussion about the Newtonian limit. For discussions of constraints derived using solar system experiments see, apart from \cite{olmo,olmo2}, also \cite{alle,cap,cap2}.

\section{\label{sec:3}Metric formalism}

As already mentioned in the introduction, the Palatini and the metric formalism give rise 
to different theories.
Starting with the same action, (\ref{action}), we get the following field equations within the metric approach:
\be
\label{genein}
f'(R) R_{\mu\nu}-\frac{1}{2}f(R)g_{\mu\nu}-\nabla_\mu\nabla_\nu f'(R)+g_{\mu\nu}\nabla^2 f'(R)=\kappa T_{\mu\nu},
\ee
where $\nabla_{\mu}$ is the covariant derivative, and $\nabla^2=\nabla^{\mu}\nabla_{\mu}$. 
Contracting we get
\be
\label{cont}
f'(R) R-2f(R)+3\nabla^2 f'(R)=\kappa T_{\mu\nu}.
\ee
For the CDTT Lagrangian, (\ref{cdtt}), we get from (\ref{cont})
\be
\label{dyn}
-R+\frac{3 a^2}{R}-\frac{6 a^2}{R^3}\nabla^2 R+\frac{18 a^2}{R^4}\nabla^{\mu}R\nabla_{\mu}R=\kappa T.
\ee
The above equation relates the scalar curvature with the stress energy tensor in a dynamical way. This makes
any analysis a lot more complicated than in the Palatini approach, where eq.~(\ref{struct}) was just an
algebraic equation.

In \cite{dick} Dick expanded around the de Sitter solution to derive the Newtonian limit in fourth-order gravity. 
In order to do so, he used a Taylor expansion for $f(R)$ and $f'(R)$, keeping only first order terms in
the correction to the scalar curvature $\delta R$. As we showed in the previous section, truncating higher order
terms is not something trivial when models with terms inversely proportional to $R$ are considered.
The condition to be fulfilled  is $\delta R \ll a$ for the CDTT model, which is the simplest and most typical example. Note that this condition
has to hold both in the metric and the Palatini formalism. We have already shown that this condition does not hold
in the Palatini version of the theory by using eq.~(\ref{rt}). The metric formalism is, of course,
totally different, and one would have to use eq.~(\ref{dyn}) to make predictions on the value of $\delta R$. 
The dynamical nature of this equation requires a different treatment from the 
one used in the previous section. 
What is clear, however, even without going any further in the analysis is that it not at all safe 
to assume that in all cases $\delta R \ll a$.
No proof that this condition holds is given in \cite{dick}, and therefore, one has to assume that higher order 
terms in the expansion of $f(R)$ and $f'(R)$ were actually
truncated ad hoc, which leaves doubts about whether the analysis presented there really applies in 
such models. This indicates that, the question of whether fourth-order gravity models with inverse powers
of the scalar curvature give good Newtonian limits, is still not answered, at least not rigorously. However, 
since, as already mentioned, the analysis of this problem seems to be completely different from the one we used for the Palatini 
approach, we intend to address it separately in future work.

As already mentioned, in \cite{olmo} the Newtonian limit of fourth order gravity was reconsidered using
the equivalent Brans--Dicke theory. In the metric formalism, an $f(R)$ theory corresponds to an $\omega=0$ 
Brans--Dicke theory. In this approach no expansion of $f(R)$ is formally needed. However, one has to expand
the potential of the scalar field $V(\phi)$ around $\phi_0$ to derive eq.~(11) and (12) of \cite{olmo}. $V(\phi)$ 
is given in terms of $f(R)$ by
\be
V(\phi)=R f'(R)-f(R),
\ee
(see eq.~(9) of \cite{olmo}).
Therefore, it is safe to assume that any problematic behavior in the expansion of $f(R)$ and $f'(R)$ will
be inherited by the expansion of $V(\phi)$. This indicates that our concerns, stated in the previous paragraph and involving the approach presented in \cite{dick},
remain relevant also for the approach followed in \cite{olmo}.

\section{\label{sec:4}Conclusions}
The Palatini version of modified models of gravity including higher order terms in the scalar curvature
has been studied in a nearly Newtonian regime. This has been done mainly, using as an example the characteristic
model of \cite{vollick}, but the results obtained were shown to be general. It has been shown that such
models can give a good Newtonian limit, as long as the coefficients of the higher order terms are reasonably
small. Additionally, some of the present results seem to indicate that there is doubt on whether  
the approach used in \cite{dick} and involved first order expansions with respect to the scalar curvature, or equivalently
 the approach presented in \cite{olmo}, are adequate to study the Newtonian limit of models with terms inversely proportional to the
scalar curvature in the gravitational action,
 in the purely metric approach. 
\section*{Acknowledgements}
The author would like to thank Stefano Liberati and John Miller for valuable discussions and comments.

\end{document}